# MAXIMUM LIKELIHOOD DECODING OF CONVOLUTIONALLY CODED NONCOHERENT ASK SIGNALS IN AWGN CHANNELS

ARAFAT AL-DWEIK, Efficient Channel Coding, Inc., Cleveland, OH, USA and FUQIN XIONG, Cleveland State University, Cleveland, OH, USA

## ABSTRACT

In this work we develop the maximum likelihood detection (MLD) algorithm for noncoherent amplitude shift keying (NCASK) systems in additive white Gaussian noise (AWGN) channels. The developed algorithm was used to investigate the performance of the NCASK system with convolutional coding and soft-decision Viterbi decoding. Tight and simple upper bounds have been derived to describe the system performance; simulation results have shown that the derived upper bounds are within 0.1 dB of the simulated points.

## 1. INTRODUCTION

Amplitude shift keying is a linear modulation technique with nonconstant envelope, the modulation is achieved by switching the carrier amplitude between two levels (0, $a$). The ASK signals can be demodulated using coherent or noncoherent demodulation, in this work noncoherent demodulation will be considered. Poor power efficiency was the main source that degrades the ASK system performance. Lately, a new system was proposed by [1, 2] in which ASK is used in orthogonal frequency division multiplexing systems (OFDM). The OFDM-ASK combination has improved the power efficiency of ASK significantly. The noncoherent OFDM-ASK system has shown significant robustness against offsets in timing, phase and frequency [2], which is not the case for coherent systems. Hence, the OFDM-ASK can be an attractive modulation scheme in the case where coherent demodulation is not feasible, and can be an efficient alternative for the conventional noncoherent frequency shift keying (FSK) systems [1].

The ASK signal can be expressed as

$$r(t) = a \cdot b_i(t) \cos(\omega t + \phi) + n(t) \quad (1)$$

where $a$ is the carrier amplitude, $\omega$ is the carrier radian frequency, $\phi$ is the carrier phase, $n(t)$ is (AWGN) with two sided power spectral density $N_0/2$, and $b(t)$ is the pulse shaping used, $i \in \{1, 2\}$. In this work, rectangular pulse shaping will be considered; therefore, $b_i(t) \in \{1, 0\}$.

The optimum receiver for signals with unknown phase can be implemented as an envelope detector or a quadrature receiver [4]. The quadrature receiver consists of two correlators; I and Q, Figure 1. However, the hardware is configured as an energy detector.

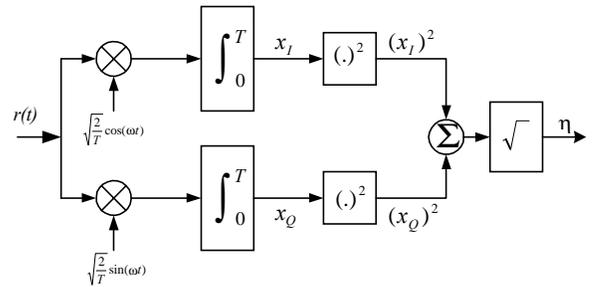

Figure 1. Noncoherent ASK quadrature receiver.

When "0" is transmitted, the output of the correlators, $x_I$ and $x_Q$, are independent Gaussian random variables with zero mean and variance $N_0/2$. The envelope ($\eta$) is defined as $\eta = \sqrt{x_I^2 + x_Q^2}$. Hence, the pdf of $\eta$ is Rayleigh [4],

$$P(\eta \mid 0) = \frac{\eta}{\sigma^2} \exp\left(-\frac{\eta^2}{2\sigma^2}\right) \quad (2)$$

where $\sigma^2$ is the AWGN variance. When "1" is transmitted, $x_I$ and $x_Q$ are also independent Gaussian random variables, however the mean has nonzero value. Therefore, the pdf of $\eta$ is Ricean and can be expressed as

$$P(\eta \mid 1) = \frac{\eta}{\sigma^2} \exp\left(-\frac{\eta^2 + E_s}{2\sigma^2}\right) I_0\left(\frac{\eta\sqrt{E_s}}{\sigma^2}\right) \quad (3)$$

where $E_s$ is the symbol energy. The Ricean distribution is usually approximated by a Gaussian distribution to derive a closed form solution of the system bit error rate (BER) [5],

$$P_e = P_b \cong \frac{1}{2}\exp\left(\frac{-E_b}{2N_o}\right) + \frac{1}{2}Q\left(\sqrt{\frac{E_b}{N_o}}\right) \quad (4)$$

where $E_b$ is the average energy per bit and $Q(\cdot)$ is the complementary error function. The BER described by (4) will be used to calculate the BER for convolutionally coded NCASK in AWGN channels with hard-decision decoding [6].

## 2. MLD OF CONVOLUTIONALLY CODED NCASK

A maximum likelihood decoder decodes a particular received sequence $\boldsymbol{\eta}$ into the code sequence $\mathbf{x}$ which maximizes the likelihood function $P(\boldsymbol{\eta}|\mathbf{x})$. This decoder provides minimum probability of error when all transmitted codewords are equal probable. Assume that a codeword $\mathbf{x} = (x_1\, x_2 \ldots x_l)$ of length $l$ is transmitted over an AWGN memoryless channel, the symbols $x_i \in \{1, 0\}$. Since each symbol is affected independently by the Gaussian noise, the joint pdf of a sequence of $l$ symbols is the multiplication of their individual pdfs. Thus, the MLD function can be expressed as

$$P(\boldsymbol{\eta} \mid \mathbf{x}) = \prod_{i=1}^{l} P(\eta_i \mid x_i) \quad (5)$$

where $P(\eta_i \mid x_i)$ is the channel transition probability. The conditional pdf $P(\eta_i \mid x_i)$ can be expressed as

$$P(\eta_i \mid x_i) = \frac{\eta_i}{\sigma^2}\exp\left(-\frac{\eta_i^2 + x_i E_s}{2\sigma^2}\right) \cdot I_0\left(x_i \frac{\eta_i \sqrt{E_s}}{\sigma^2}\right) \quad (6)$$

Thus,

$$P(\boldsymbol{\eta} \mid \mathbf{x}) = \prod_{i=1}^{l} \frac{\eta_i}{\sigma^2}\exp\left(-\frac{\eta_i^2 + x_i E_s}{2\sigma^2}\right) \cdot I_0\left(x_i \frac{\eta_i \sqrt{E_s}}{\sigma^2}\right) \quad (7)$$

Since the natural logarithm is monotonic function, we can take the natural logarithm of both sides of (7), it is usually called the log-likelihood function (metric) of the path $\mathbf{x}$. Therefore,

$$\ln[P(\boldsymbol{\eta} \mid \mathbf{x})] = \sum_{i=1}^{l}\left[\underbrace{\ln\left(\frac{\eta_i}{\sigma^2}\right) - \frac{\eta_i^2}{2\sigma^2}}_{A} - \frac{E_s}{2\sigma^2} x_i + \ln\left(I_0\left(x_i \eta_i \frac{\sqrt{E_s}}{\sigma^2}\right)\right)\right] \quad (8)$$

The term $A$ is independent of the path selected and can be considered as a constant. Hence, the MLD for noncoherent ASK can be obtained by calculating the path metric for all possible paths and selecting the path with maximum metric.

## 3. PERFORMANCE OF NCASK WITH SOFT-DECISION DECODING

Without loss of generality, we can assume that the all zero codeword $\mathbf{X}$ is transmitted. The Hamming distance between the all zero path and any other path is $d$. The decoder calculates the metrics for all possible paths; the path with the maximum metric will be selected. Since the path with minimum $d$ ($\mathbf{Y}$) is the closest to $\mathbf{X}$, the bound will take only that path into consideration [6]. The hamming distance of $\mathbf{Y}$ is equal to $d_{free}$ and it merges with the all zero path after $l$ samples. Therefore, the decoder accumulates $l$ soft symbol metrics for the two paths, the two metrics will be compared and the path with maximum metric will be selected. The first event error probability ($P_d$) occurs when the metric of $\mathbf{Y}$ is greater than the metric of $\mathbf{X}$. Therefore,

$$P_d = P(\ln[P(\boldsymbol{\eta} \mid \mathbf{Y})]) > P(\ln[P(\boldsymbol{\eta} \mid \mathbf{X})]) \quad (9)$$

$$= P\left[\begin{array}{l} \sum_{i=1}^{l}\left[A - \frac{E_s}{2\sigma^2} y_i + \ln\left(I_0\left(y_i \eta_i \frac{\sqrt{E_s}}{\sigma^2}\right)\right)\right] > \\ \sum_{j=1}^{l}\left[A - \frac{E_s}{2\sigma^2} x_j + \ln\left(I_0\left(x_j \eta_j \frac{\sqrt{E_s}}{\sigma^2}\right)\right)\right] \end{array}\right] \quad (10)$$

Since $\mathbf{X}$ represents all-zero path, (10) can be simplified to

$$P_d = P\left[\sum_{i=1}^{l}\left[\frac{-E_s}{2\sigma^2} y_i + \ln\left(I_0\left(y_i \eta_i \frac{\sqrt{E_s}}{\sigma^2}\right)\right)\right] > 0\right] \quad (11)$$

at high signal-to-noise ratio (SNR), the second term in (11) can be simplified using the following approximation,

$$\ln\left(I_0\left(\frac{\eta\sqrt{E_s}}{\sigma^2}\right)\right) \approx \frac{\eta\sqrt{E_s}}{\sigma^2} - 1 \quad (12)$$

substituting this approximation in (11) and noting that $\mathbf{Y}$ has only $d_{free}$ 1's out of the $l$ symbols, (11) can be written as

$$P_d = P\left[\left[\sum_{i=1}^{d_{free}} \eta_i\right] > \left[d_{free}\frac{\sqrt{E_s}}{2} + \frac{\sigma^2}{\sqrt{E_s}} d_{free}\right]\right] \quad (13)$$

Since we assumed that the all-zero path was transmitted, the samples $\eta_i$ are i.i.d. random variables with Rayleigh distribution. For large $d_{free}$ ($\geq 10$), the central limit theorem can be applied and the summation term in (13) can be

approximated by a Gaussian random variable ($Z$). The mean of $Z$ can be expressed as

$$m_Z = \sum_{i=1}^{d_{free}} \sqrt{\frac{\pi}{2}}\sigma = d_{free}\sqrt{\frac{\pi}{2}}\sigma \qquad (14)$$

where $\sigma^2$ is the variance of the AWGN. The variance of $Z$ can be expressed as

$$\sigma_Z^2 = \sum_{i=1}^{d_{free}} (2-\frac{\pi}{2})\sigma^2 = d_{free}(2-\frac{\pi}{2})\sigma^2 \qquad (15)$$

The quality of this approximation was investigated using simulation, Figure 2. The simulations were carried over a wide range of $E_b/N_0$, it is clear that the approximation is independent of the SNR. Hence, the Gaussian approximation is accurate and $Z$ can be closely approximated by a Gaussian random variable with mean $m_Z$ and variance $\sigma_Z^2$.

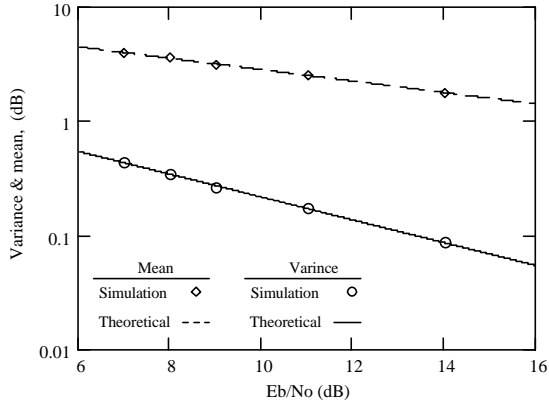

Figure 2. Theoretical and simulated mean and variance of the Gaussian approximation.

Hence, The pairwise error probability will be

$$P_d = \int_{\gamma}^{\infty} \frac{1}{\sqrt{2\pi}\sigma_Z} \exp\left(-\frac{(x-m_Z)^2}{2\sigma_Z^2}\right)dx \qquad (16)$$

The threshold $\gamma$ can be expressed as

$$\gamma = d_{free}\left(\frac{\sqrt{E_s}}{2}+\frac{\sigma^2}{\sqrt{E_s}}\right) \qquad (17)$$

By evaluating the integration in (16), $P_d$ can be expressed as

$$P_d = \frac{1}{2} - \frac{1}{2}\mathrm{erf}\left[\frac{\sqrt{d_{free}}}{\sqrt{8-2\pi}}\left(\sqrt{\frac{E_s}{N_0}}+\frac{1}{\sqrt{E_s/N_0}}-\sqrt{\pi}\right)\right] \qquad (18)$$

where $\mathrm{erf}(\cdot)$ is the error function. Following the same procedure used in [6], the bit error probability can be upper bounded by

$$P_b \leq \frac{1}{k}B_{dfree}\left[\frac{1}{2}-\frac{1}{2}\mathrm{erf}\left[\frac{\sqrt{d_{free}}}{\sqrt{8-2\pi}}\left(\sqrt{\frac{E_s}{N_0}}+\sqrt{\frac{N_0}{E_s}}-\sqrt{\pi}\right)\right]\right] \qquad (19)$$

where $B_{dfree}$ is the number of non-zero bits on all weight $d_{free}$ paths. In terms of $E_b/N_0$, the BER can be upper bounded by

$$P_b \leq$$
$$\frac{1}{2k}B_{dfree}\left[1-\mathrm{erf}\left[\frac{\sqrt{d_{free}}}{\sqrt{8-2\pi}}\left(\sqrt{2R\frac{E_b}{N_0}}+\sqrt{\frac{N_0}{2R\cdot E_b}}-\sqrt{\pi}\right)\right]\right] \qquad (20)$$

For binary ASK signals $E_s = 2R\cdot E_b$.

To verify the tightness of the derived bound; the system was simulated for two different codes. The first code has $d_{free}=10$, the code rate ($R$) is equal to 1/2, and the constraint length ($K$) is equal to 7, Figure 3.

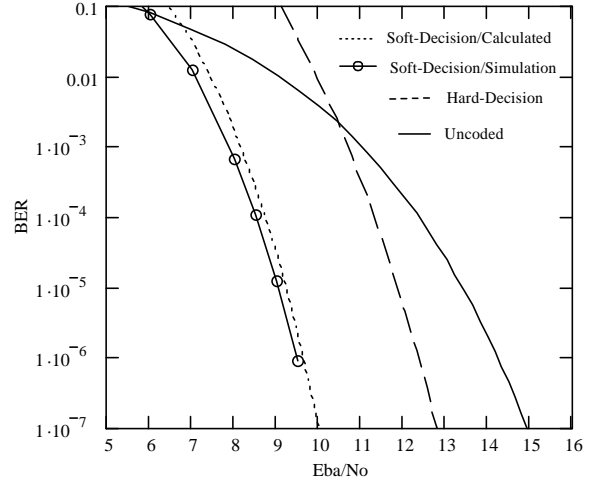

Figure 3. Coded system performance with soft decoding, $d_{free}=10$.

The second code has $R=1/3$, $d_{free}=15$ and $K=7$, Figure 4. At high SNR, the simulation points for $d_{free}=10$ are within 0.1 dB from the upper bound. For $d_{free}=15$, the simulation points are within 0.05 dB.

The curves for the hard-decision decoding in Figure 3 and Figure 4 were calculated using the following [6],

$$P_b \approx \frac{1}{k}B_{dfree}\cdot 2^{d_{free}}\cdot p^{d_{free}/2} \qquad (21)$$

where $p$ is given in (4). It can be seen that soft-decision decoding has improved the system BER performance by approximately 3 dB when compared to hard-decision decoding.

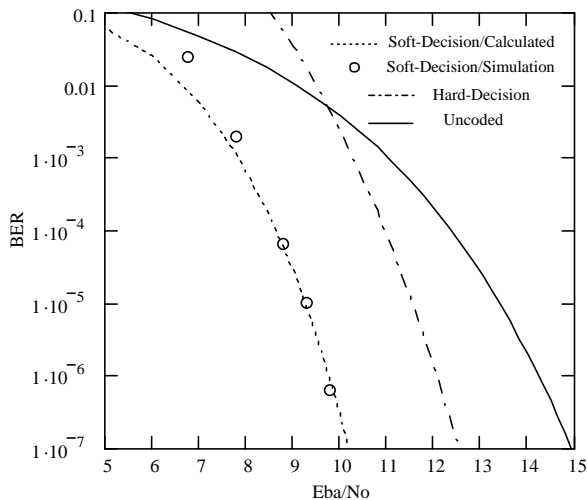

Figure 4. Coded system performance with soft decoding, $d_{free}$=15.

## 4. RESULTS AND CONCLUSION

In this paper, the MLD of convolutionally coded noncoherent ASK signals was derived. The system performance was investigated using the developed MLD algorithm. The derived upper bounds were within 0.1 dB from the simulation points at $E_b/N_0$ of 10 dB and $d_{free}$=10, and within 0.05 dB for $d_{free}$=15. The accuracy of the Gaussian approximation was tested using simulation; the theoretical mean and variance were compared to the simulated ones for several values of $E_b/N_0$. The difference between the calculated and the simulated points has never exceeded 0.05% for all values of $E_b/N_0$.